\documentclass{article}

\usepackage{arxiv}
\usepackage{amsmath}
\usepackage{amssymb}
\usepackage{amsthm}
\usepackage[utf8]{inputenc} 
\usepackage[T1]{fontenc}    
\usepackage{hyperref}
\usepackage{url}            
\usepackage{booktabs}       
\usepackage{amsfonts}       
\usepackage{nicefrac}       
\usepackage{microtype}      
\usepackage{lipsum}		
\usepackage{graphicx}
\usepackage{doi}

\usepackage{mathtools} 
\usepackage{float}

\title{Diatomic Molecules in deSitter and Anti-deSitter Spaces}

\author{ \href{https://orcid.org/0000-0002-8191-9697}{\includegraphics[scale=0.06]{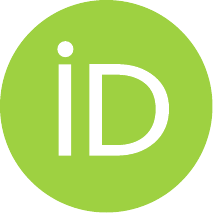}\hspace{1mm}Meriem Abdelaziz}\thanks{Use footnote for providing further
		information about author (webpage, alternative
		address)---\emph{not} for acknowledging funding agencies.} \\
	Laboratory PPNM, Department of Matter Sciences\\
	University of Biskra, 07000 Algeria \\
	\texttt{meriem.abdelaziz@univ-biskra.dz} \\
		\And
	\href{https://orcid.org/0000-0002-8096-6280}{\includegraphics[scale=0.06]{orcid.pdf}\hspace{1mm}Mustafa Moumni} \\
	LPRIM, Department of Physics\\
	University of Batna 1, 05000 Algeria \\
	\texttt{m.moumni@univ-batna.dz} \\
 \And
	\href{https://orcid.org/0000-0002-0466-9559}{\includegraphics[scale=0.06]{orcid.pdf}\hspace{1mm}Mokhtar Falek} \\
	LPPNM, Department of Matter Sciences\\
	University of Biskra, 07000 Algeria \\
	\texttt{mokhtar.falek@univ-biskra.dz} \\
	}

\renewcommand{\shorttitle}{Diatomic Molecules in deSitter and Anti-deSitter Spaces}

\hypersetup{
pdftitle={Diatomic Molecules in deSitter and Anti-deSitter Spaces},
pdfsubject={q-bio.NC, q-bio.QM},
pdfauthor={M. AbdelAziz, M. Moumni and M. Falek},
pdfkeywords={diatomic molecules, (anti-)deSitter, Nikiforov-Uvarov method},
}

\begin{document}
\maketitle

\begin{abstract}
The Schr\"{o}dinger equation for diatomic molecules in deSitter and anti-deSitter spaces is studied using the extended uncertainty principle formulation. The equations are solved by the Nikiforov-Uvarov method for both the Kratzer potential and the pseudoharmonic oscillator. The energy
eigenvalues of the system have been derived analytically, and the exact expressions of the eigenfunctions are provided in terms of Romanovski and Jacobi polynomials. The impact of the spatial deformation parameter on the bound states is also examined, with experimental results used to establish an upper limit for this parameter.
\end{abstract}


\section{Introduction}
The extension of the quantum field theory to curved spacetime, which can be seen as a first approximation of quantum gravity, has generated significant interest due to numerous motivations aiming to regulate infinities in standard field theories \cite{Hossenfelder2013} \cite{Tawfik2015} \cite{Frassino2018} \cite{Todorinov2021} \cite{Bosso2023} (and the references therein). In the case of such a curved spacetime, we are faced with a structure disturbed by the gravitational field. These modifications can also be found in the Snyder model, where measurements in non-commutative quantum mechanics can be governed by a Generalized Uncertainty Principle (GUP) \cite{Snyder1947} \cite{Hellund1954} \cite{Kadyshevskii1963} \cite{Hamil2018} \cite{Meljanac2023}. This model exhibits a fundamental length scale of the order of the Planck length, resulting in a non-negligible minimal uncertainty in position measurement \cite{Golfand1960} \cite{Hadj2018} \cite{Mignemi2010}. Given that multiple arguments suggest that quantum gravity also implies a measurable minimal length of the order of the Planck length, considerable efforts have been made to extend the study of quantum mechanics to curved spacetime using the Extended Uncertainty Principle (EUP) \cite{Mignemi2010} \cite{Tawfik2015} (and the references therein). A significant consequence arising from these extensions is that the minimal measurable length in quantum gravity can be linked to a modification of the standard Heisenberg algebra by introducing small corrections to canonical commutation relations \cite{Hossenfelder2013} \cite{Frassino2018} \cite{Bosso2023}. These modifications find their justification in various recent theories such as string theory \cite{Konishi1990}, black hole physics \cite{Scardigli1999}, non-commutative geometry \cite{Douglas2001}, special double relativity (DSR) \cite{Amelino2001}, extra dimensions theories \cite{Scardigli2003} and even in the study of the effects of
Newtonian gravity on quantum systems \cite{Kuzmichev2020}.

Although quantum gravity models remain primarily theoretical, the phenomenological analysis of their various aspects is constantly expanding. As a result, a large number of studies are currently addressing the "phenomenology of quantum gravity" \cite{Amelino2002} \cite{Liberati2011} 
\cite{FaragAli2011} \cite{Riasat2023} \cite{Bevilacqua2023}. This realm of investigation encompasses the impact of deformed uncertainty principles on the solutions of wave equations. For relativistic quantum mechanics, the list of exactly solvable problems is very restricted. For instance, a recent study examined the cases of one-dimensional Dirac and Klein-Gordon
oscillators in anti-de Sitter (AdS) space \cite{Hamil2018}, as well as the two-dimensional and three-dimensional Dirac oscillators in the presence of minimal momentum uncertainty \cite{Hadj2018}. Additionally, an exact solution of the (1 + 1)-dimensional bosonic oscillator under the influence
of a uniform electric field in AdS space has been studied \cite{Hamil2019a}.

On the other hand, the non-relativistic case also presents great interest but remains unexplored in this framework. Although it is not possible to derive a non-relativistic Schr\"{o}dinger-like covariant equation from the covariant Klein-Fock-Gordon equation in the conventional field theory approach, we can use the EUP formulation to obtain the dS and AdS versions of the Schr\"{o}dinger equation. Indeed, Hamil et al. dealt with the exact solution of the Schr\"{o}dinger equation in D dimensions for the free particle and the harmonic oscillator in AdS space \cite{Hamil2019b}. In addition, Chung analytically studied the one-dimensional box problem and the harmonic oscillator problem \cite{Chung2019}. We also find the exact solutions for the hydrogen atom in the two cases dS at AdS \cite{Falek2020}. However, further investigations are still needed to explore the non-relativistic case in this framework.

This study investigates the deformed Schr\"{o}dinger equation using the EUP formulation for both the pseudoharmonic oscillator (PHO) and the Kratzer potentials. The PHO potential was first studied by Gol'dman et al. in 1960 as a description of the rotational and the vibrational states of diatomic molecules \cite{Goldman1960}. Due to its importance in the fields of chemical and molecular physics, a number of authors have carried out extensive research on this topic \cite{Dong2005} \cite{Dong2007} \cite{Oyewumi2008} \cite{Oyewumi2012} \cite{Fernandez2024}. Moreover, this potential is often used as confinement potential in quantum dots (QDs) \cite{Liang2019} \cite{Baazouzi2020} \cite{Kenfack2024}. Its application is crucial for the theoretical description of the electronic properties of QDs
and for the manufacture of nanoscale devices \cite{Chakraborty1999} \cite{Pal2019} \cite{Bejan2019}. On the other hand, the Kratzer potential was originally developed to investigate patterns in the band spectra of diatomic molecules \cite{Kratzer1920a} \cite{Kratzer1920b}. Since then, it has garnered considerable attention for its usefulness in a  variety of fields, including nuclear physics  \cite{Fortunato2003}, quantum chemistry \cite{Berkdemir2005}, molecular physics \cite{Hajigeorgiou2006} and chemical physics \cite{Van2006}. Recently, the Kratzer potential is employed to
examine semiconductor quantum dots and their optical properties \cite{Batra2018} \cite{Heddar2019} \cite{Jaouane2023}.

The article is structured as follows: following this initial section serving as the introduction, the second section \ref{sec:II} reviews deformed quantum mechanics relationships. The third section \ref{sec:III} explains the Nikiforov-Uvarov method in detail, while the fourth section \ref{sec:IV} presents explicit calculations related to
the Schr\"{o}dinger equation applied to diatomic  molecules in
(anti-)de Sitter space, considering pseudo-harmonic and Kratzer potentials. Finally, the fifth section \ref{sec:V} provides a detailed analysis, and the sixth section \ref{sec:VI} presents the article's findings.

\section{Review of the deformed quantum mechanic relations}
\label{sec:II}
Commutation relations resulting in EUP in three-dimensional space are defined by a deformed Heisenberg algebra \cite{Mignemi2012} \cite{Stetsko2012}. 
\begin{equation}
\left[ X_{i},X_{j}\right] =0,\text{ }\left[ P_{i},P_{j}\right] =i\hbar
\kappa \lambda \epsilon _{ijk}L_{k},\text{ }\left[ X_{i},P_{j}\right]
=i\hbar \left( \delta _{ij}-\kappa \lambda X_{i}X_{j}\right) ,\text{ }\kappa
=-1,+1  \label{1}
\end{equation}
where $\lambda $ is a small positive parameter. In the context of quantum gravity, this EUP parameter $\lambda $ is determined as the fundamental constant associated with the scale factor of the expanding universe. It is proportional to the cosmological constant $\Gamma $ which is calculated as $
\Gamma =-3\lambda =-3a^{-2}$ with $a$ being the anti-de Sitter radius \cite{Bolen2005}. The components of angular momentum $L_{k}$ are expressed as follows: 
\begin{equation}
L_{k}=\epsilon _{ijk}X_{i}P_{j},  \label{2}
\end{equation}%
satisfying the standard algebra:
\begin{equation}
\left[ L_{i},P_{j}\right] =i\hbar \varepsilon _{ijk}P_{k},\text{ }\left[
L_{i},X_{j}\right] =i\hbar \varepsilon _{ijk}X_{k},\text{ }\left[ L_{i},L_{j}%
\right] =i\hbar \varepsilon _{ijk}L_{k}  \label{3}
\end{equation}
As in ordinary quantum mechanics, the commutation relations in eq.(1) lead to Heisenberg uncertainty relations:
\begin{equation}
\Delta X_{i}\Delta P_{i}\geq \frac{\hbar }{2}\left( 1+\lambda \left( \Delta
X_{i}\right) ^{2}\right)   \label{4}
\end{equation}
where we choose the states for which $\left\langle X_{i}\right\rangle =0$.

Two types of subalgebra are distinguished based on the value of $\kappa $.

When $\kappa =-1$, the deformed algebra is characterized by the presence of a nonzero minimum uncertainty in momentum, and it is called the Anti-deSitter (AdS) model. For simplicity, we assume isotropic uncertainties $X_{i}=X$, which allows us to write the minimal uncertainty for the momentum in the AdS model:
\begin{equation}
\left( \Delta P_{i}\right) _{\min }=\hbar \kappa \sqrt{\lambda }  \label{5}
\end{equation}
For $\kappa =+1$, the deSitter (dS) model is applicable and eq.(4) does not imply a non-zero minimum value for the uncertainties of the momentum.
This can be seen in Fig.1, where the uncertainty relations are plotted in accordance with the modified relation of eq.(4). The forbidden region for position and momentum measurements in AdS space is represented by the coloured region in the figure.
\begin{figure} [htb]
    \centering
    \includegraphics[scale=0.7]{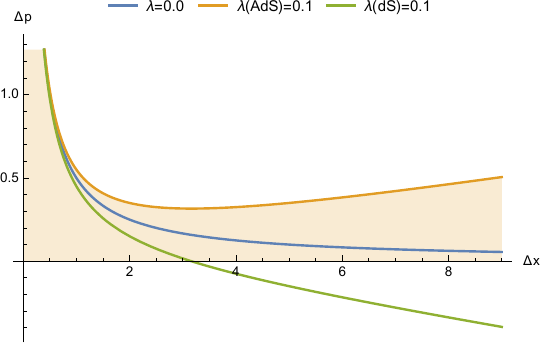}
    \caption{HUP vs EUP for both dS and AdS spaces}
    \label{fig:1}
\end{figure}
The operators $X_{i}$ and $P_{i}$ satisfy the modified algebra in eq.(1), which gives rise to the rescaled uncertainty relation in momentum space in eq.(4). To study the exact solutions of the deformed Schr\"{o}dinger equation, we represent these operators as functions of the operators $x_{i}$ and $p_{i} $ that satisfy the ordinary canonical commutation relations:
\begin{subequations}
\label{6}
\begin{align}
X_{i}& =\frac{x_{i}}{\sqrt{1+\kappa \lambda r^{2}}}  \label{6a} \\
P_{i}& =-i\hbar \sqrt{1+\kappa \lambda r^{2}}\partial _{x_{i}}  \label{6b}
\end{align}
\end{subequations}
The variable $r$ is defined within the specified domain $\left] -1/\sqrt{\lambda },1/\sqrt{\lambda }\right[ .$

\section{The Nikiforov--Uvarov method}
\label{sec:III}
The NU-method transforms second-order differential equations (ODEs) into the hypergeometric type through an appropriate coordinate transformation. Initially, it is necessary to transform the ODE into the following form:
\begin{equation}
\psi^{\prime \prime }\left( s\right) +\frac{\tilde{\tau}\left(
s\right) }{\sigma \left( s\right) }\psi ^{\prime }\left( s\right) +\frac{
\tau \left( s\right) }{\tilde{\sigma}\left( s\right) }\psi \left( s\right) =0
\label{7}
\end{equation}
where $\sigma \left( s\right) $ and $\widetilde{\sigma }\left( s\right) $ are polynomials with at most second degree, and $\widetilde{\tau }\left(\right) $ is a polynomial with at most first degree \cite{Nikiforov1988}.
If we consider the following factorization:
\begin{equation}
\psi \left( s\right) =\phi \left( s\right) y\left( s\right)  \label{8}
\end{equation}
\begin{equation}
\sigma \left( s\right) y^{\prime \prime }\left( s\right) +\tau
\left( s\right) y^{\prime }\left( s\right) +\Lambda y\left( s\right) =0
\label{9}
\end{equation}
where:
\begin{equation}
\tau \left( s\right) =\tilde{\tau}\left( s\right) +2\pi \left( s\right) 
\text{\ and }\pi \left( s\right) =\sigma \left( s\right) \frac{d}{ds}\left(
\ln \phi \left( s\right) \right) \text{\ }  \label{10}
\end{equation}%
the $\Lambda $ will be defined by:
\begin{equation}
\Lambda _{n}+n\tau ^{\prime }+\frac{n\left( n-1\right) \sigma
^{\prime \prime }}{2}=0,\text{ \ \ }n=0,1,2,...  \label{11}
\end{equation}
The energy eigenvalues are calculated from the equation above. To do so, it is necessary to determine $\pi \left( s\right) $ and $\Lambda $ by defining:
\begin{equation}
k=\Lambda -\pi ^{\prime }\left( s\right)  \label{12}
\end{equation}%
Solving this quadratic equation for $\pi \left( s\right) $, we get:
\begin{equation}
\pi \left( s\right) =\left( \frac{\sigma ^{\prime }-\widetilde{\tau }}{2}%
\right) \pm \sqrt{\left( \frac{\sigma ^{\prime }-\tilde{\tau}}{2}\right)
^{2}-\tilde{\sigma}+\sigma k}  \label{13}
\end{equation}%
Here $\pi \left( s\right) $ is represented by a polynomial with the parameter $s$, and the prime factors indicate the derivatives of the first order.

The determination of $k$ is the essential point in the calculation of $\pi\left( s\right) $. It is simply defined so that the discriminant of the square root must be zero. This gives us a general quadratic equation for $k$. The wave function follows naturally, using eq.(10) and the Rodrigues relation:%
\begin{equation}
y_{n}\left( s\right) =\frac{C_{n}}{\rho \left( s\right) }\frac{d^{n}}{ds^{n}}%
\left[ \sigma ^{n}\left( s\right) \rho \left( s\right) \right]  \label{14}
\end{equation}%
where $C_{n}$ is the normalizing constant and the weight function $\rho\left( s\right) $ obeys the following relationship:%
\begin{equation}
\frac{d}{ds}\left[ \sigma \left( s\right) \rho \left( s\right) \right] =\tau
\left( s\right) \rho \left( s\right)  \label{15}
\end{equation}%
This equation refers to the classical orthogonal polynomials and the relation of orthogonality can be defined as follows:%
\begin{equation}
\int_{a}^{b}y_{n}\left( s\right) y_{m}\left( s\right) \rho \left( s\right)
ds=0,\text{ \ }m\neq n  \label{16}
\end{equation}

\section{Schr\"{o}dinger equation for diatomic molecules in (anti-)de Sitter spaces}
\label{sec:IV}
In this section we study the effects of the deformed space on the energy eigenvalues and eigenfunctions in the context of non-relativistic quantum mechanics. In three dimensions, the Schr\"{o}dinger equation for a central interaction is as follows:%
\begin{equation}
\left[ \frac{\mathbf{p}^{2}}{2m}+V\left( r\right) \right] \psi \left( 
\mathbf{r}\right) =E\psi \left( \mathbf{r}\right) \,  \label{17}
\end{equation}

To include the effect of the EUP on the above Schr\"{o}dinger equation, we use the first order transformations (6.a) and (6.b). To uncouple the equation according to the symmetry of the potential, we write the solution as $\psi \left( r,\theta ,\varphi \right) =R\left( r\right) Y_{l}^{m_{l}}\left( \theta ,\varphi \right) $, which allows us to split the equation into two parts, one angular and the other radial:%
\begin{equation}
L^{2}Y_{l}^{m_{l}}\left( \theta ,\varphi \right) =\hbar ^{2}l\left(
l+1\right) Y_{l}^{m_{l}}\left( \theta ,\varphi \right)   \label{18}
\end{equation}%
\begin{equation}
\left[ -\frac{\hbar ^{2}}{2m}\left( \left( 1+\kappa \lambda r^{2}\right)
\left( \frac{d^{2}}{dr^{2}}+\frac{2}{r}\frac{d}{dr}-\frac{l\left( l+1\right) 
}{r^{2}}\right) +\kappa \lambda r\frac{d}{dr}\right) + 
V\left( r\right) -\frac{\kappa \lambda }{2}r^{3}\frac{dV\left(
r\right) }{dr}\right] R\left( r\right) =ER\left( r\right)   \label{19}
\end{equation}%
We are interested in the resolution of the radial equation because the angular equation of the system is just the usual one for spherical harmonics.

\subsection{The Pseudo-Harmonic Potential}
In this case, the PHO potential take the form \cite{Dong2003} \cite{Sever2007}: 
\begin{equation}
V\left( r\right) =D_{e}\left( \frac{r}{r_{e}}-\frac{r_{e}}{r}\right) ^{2}
\label{20}
\end{equation}%
where $D_{e}$ is the dissociation energy between two atoms in a solid and $r_{e}$ is the equilibrium inter-molecular spacing. So the expression of eq.(19) will be as follows:%
\begin{equation}
\left[ \left( y\frac{d}{dr}\right) ^{2}+\frac{2y^{2}}{r}\frac{d}{dr}-\frac{%
\delta y^{2}}{r^{2}}-\frac{\eta r^{2}}{y^{2}}+\varepsilon \right] U\left(
r\right) =0  \label{21}
\end{equation}%
where we have used the following ansatz $R\left( r\right) =U(r)/\sqrt{r}$, the variable $y\equiv \sqrt{1+\kappa \lambda r^{2}}$ and the notations: 
\begin{equation}
\delta =l\left( l+1\right) +\frac{2mD_{e}r_{e}^{2}}{\hbar ^{2}},\eta =\frac{%
2m}{\hbar ^{2}}\frac{D_{e}}{r_{e}^{2}},\varepsilon =\frac{2m}{\hbar ^{2}}%
\left( E+2D_{e}\right)  \label{22}
\end{equation}%
To obtain the exact solution of eq.(21), we use the following transformations $F\left( r\right) =y^{\mu }g\left(y\right) $ with the new variable $y$. This gives a new form for the radial equation:
\begin{equation}
\left[ \left( 1-y^{2}\right) \frac{d^{2}}{dy^{2}}+\left( \frac{2\mu }{y}%
-\left( 2\mu +3\right) y\right) \frac{d}{dy}-\frac{\delta y^{2}}{1-y^{2}}+%
\frac{\varepsilon }{\kappa \lambda }-3\mu \right] g\left( y\right) =0
\label{23}
\end{equation}%
To get this equation above, the free parameter $\mu $ is chosen so that it verifies the relation:
\begin{equation}
\mu \left( \mu -1\right) -\frac{\eta }{\lambda ^{2}}=0  \label{24}
\end{equation}%
This equation gives two possible solutions:
\begin{equation}
\mu _{1,2}=\frac{1}{2}\left( 1\pm \sqrt{1+\frac{8mD_{e}}{\lambda ^{2}\hbar
^{2}r_{e}^{2}}}\right)  \label{25}
\end{equation}%
From the expression of $F(r)$, the function $g(y)$ should be nonsingular at $y=\pm 1$; So, the accepted value of $\mu $ is:%
\begin{equation}
\mu =\frac{1}{2}\left( 1+\sqrt{1+\frac{8mD_{e}}{\lambda ^{2}\hbar
^{2}r_{e}^{2}}}\right)  \label{26}
\end{equation}

Now, using another change of the variable $s=2y^{2}-1$ in eq.(23), we reduce this equation to a class of known differential equations with a polynomial solution and obtain the following form:%
\begin{equation}
\left[ \left( 1-s^{2}\right) ^{2}\dfrac{d^{2}}{ds^{2}}+\left( 1-s^{2}\right)
\left( \left( \mu -1\right) -\left( \mu +2\right) s\right) \dfrac{d}{ds}%
+a_{1}s^{2}+a_{2}s+a_{3}\right] g\left( s\right) =0  \label{27}
\end{equation}%
where we have:%
\begin{equation}
a_{1,3}=\frac{-1}{4}\left( \delta \pm \frac{\varepsilon }{\kappa \lambda }%
\mp 3\mu \right) ,\text{ }a_{2}=\frac{-\delta }{2}  \label{28}
\end{equation}%
The comparison between eq.(27) and eq.(7) allows us to use the NU method, where the expressions of the polynomials that appear in eq.(7) are given by:%
\begin{equation}
\sigma \left( s\right) =\left( 1-s^{2}\right) \text{ , }\tilde{\tau}\left(
s\right) =\left( \mu -1\right) -\left( \mu +2\right) s\text{ \ and }\tilde{%
\sigma}\left( s\right) =a_{1}s^{2}+a_{2}s+a_{3}  \label{29}
\end{equation}%
If we insert them in eq.(13), we get:%
\begin{equation}
\pi \left( s\right) =\pm \sqrt{\left( \frac{\mu ^{2}}{4}-a_{1}-k\right)
s^{2}-\left( \frac{\mu \left( \mu -1\right) }{2}+a_{2}\right) s+\frac{\left(
\mu -1\right) ^{2}}{4}-a_{3}+k}  +\frac{\left( \mu s-\left( \mu -1\right) \right) }{2}
\label{30}
\end{equation}
The parameter $k$ is determined as described in the previous section; We obtain two values:%
\begin{equation}
k_{1,2}=\frac{1}{4}\left[ -\frac{\varepsilon }{\tau \lambda }-3\mu +\left(
\mu -\frac{1}{2}\right) \left( 1\pm 2\sqrt{\delta +\frac{1}{4}}\right) %
\right]  \label{31}
\end{equation}

For $\pi (s)$, we get the following possible solutions:%
\begin{equation}
\pi \left( s\right) =\left\{ 
\begin{array}{c}
\pi _{1,3}=\frac{1}{2}\left( \left( 2\mu \mp \sqrt{\delta +\frac{1}{4}}-%
\frac{1}{2}\right) s\mp \sqrt{\delta +\frac{1}{4}}+\frac{3}{2}-2\mu \right) 
\text{ } \\ 
\pi _{2,4}=\frac{1}{2}\left( \left( \pm \sqrt{\delta +\frac{1}{4}}+\frac{1}{2%
}\right) \left( s+1\right) \right)%
\end{array}%
\right.  \label{32}
\end{equation}%
where $\pi _{1}$ and $\pi _{2}$ relate to $k_{1}$ while $\pi _{3}$ and $\pi_{4}$ relate to $k_{2}$; The correct solution is $\pi _{4}$, so:
\begin{equation}
\tau (s)=-\left( \mu +\sqrt{\delta +\frac{1}{4}}+\frac{3}{2}\right) s+\left(
-\mu +\sqrt{\delta +\frac{1}{4}}+\frac{1}{2}\right)   \label{33}
\end{equation}%
We calculate $\Lambda $\ from eq.(11):%
\begin{equation}
\Lambda =k_{2}+\frac{1}{2}\left( -\sqrt{\delta +\frac{1}{4}}+\frac{1}{2}%
\right) =n\left( n+\mu +\sqrt{\delta +\frac{1}{4}}-\frac{1}{2}\right) ,n\in \mathbb{N}
\label{34}
\end{equation}%
and the energy eigenvalues follows:%
\begin{multline}
E_{n,l,\kappa }=\hbar \sqrt{\frac{2D_{e}}{mr_{e}^{2}}}\sqrt{1+\frac{\lambda
^{2}\hbar ^{2}r_{e}^{2}}{8mD_{e}}}\left( 2n+\sqrt{\left( l+\frac{1}{2}%
\right) ^{2}+\frac{2mD_{e}r_{e}^{2}}{\hbar ^{2}}}+1\right) -  \label{35} \\
\frac{\lambda \kappa \hbar ^{2}}{m}\left( (n+\frac{1}{2})\left( 2n+2\sqrt{%
\left( l+\frac{1}{2}\right) ^{2}+\frac{2mD_{e}r_{e}^{2}}{\hbar ^{2}}}%
+1\right) -\frac{1}{4}\right) -2D_{e}
\end{multline}

It is observed that the contribution of the deformation to energies is not uniform and exhibits variations from one level to another. These disparities may even lead to a reversal in the order of levels for extreme values of the parameter $\lambda $.
Let us now find the corresponding eigenfunctions. Taking the expression of $\pi _{4}\left( s\right) $ from eq.(34), the $\phi \left( s\right) $ part is defined from the relation eq.(10): 
\begin{equation}
\phi \left( s\right) =\left( 1+s\right) ^{-\frac{\mu }{2}+\frac{1}{2}\sqrt{%
\delta +\frac{1}{4}}+\frac{1}{4}}\left( 1-s\right) ^{\frac{1}{2}\left( \mu
-1\right) }  \label{36}
\end{equation}%
and according to the form of $\sigma \left( s\right) $ in eq.(15), the $y\left( s\right) $ part is given by the Rodrigues relation: 
\begin{equation}
y_{n}\left( s\right) =\frac{C_{n}}{\rho \left( s\right) }\frac{d^{n}}{ds^{n}}%
\left[ \left( 1-s^{2}\right) ^{n}\rho \left( s\right) \right]  \label{37}
\end{equation}%
where $\rho \left( s\right) =\left( 1+s\right) ^{\sqrt{\delta +\frac{1}{4}}}\left( 1-s\right) ^{\left( \mu -\frac{1}{2}\right) }$. This expression stands for the Jacobi polynomials, so:%
\begin{equation}
y_{n}\left( s\right) \equiv P_{n}^{\left( \mu -\frac{1}{2},\sqrt{\delta +%
\frac{1}{4}}\right) }\left( s\right)  \label{38}
\end{equation}%
Hence, $g(s)$ can be written in the following form:%
\begin{equation}
g(s)=C_{n}\left( 1+s\right) ^{-\frac{\mu }{2}+\frac{1}{2}\sqrt{\delta +\frac{%
1}{4}}+\frac{1}{4}}\left( 1-s\right) ^{\frac{1}{2}\left( \mu -1\right)
}P_{n}^{\left( \mu -\frac{1}{2},\sqrt{\delta +\frac{1}{4}}\right) }\left(
s\right)  \label{39}
\end{equation}%
where $C_{n}$ is a normalization constant.

Using the variable $r$, we can now write the general form of the wavefunction $\psi $ as follows:

\begin{multline}
    \psi_{n,\tau }\left( r,\theta ,\varphi \right)=C_{n}2^{\frac{1}{2}\sqrt{\delta +\frac{1}{4}}+\frac{1}{4}}\left( 1+\kappa \lambda r^{2}\right) ^{-\frac{\mu }{2}+\frac{1}{2}\sqrt{\delta +\frac{1}{4}}+\frac{1}{4}}\left(-\kappa \lambda r^{2}\right) ^{\frac{1}{2}\left( \mu -1\right)}   \label{40} \\
 \times P_{n}^{\left( \mu -\frac{1}{2},\sqrt{\delta +\frac{1}{4}}\right)}\left( 1+2\kappa \lambda r^{2}\right) Y_{l}^{m_{l}}\left( \theta ,\varphi \right)
\end{multline}

\subsection{The Kratzer Potential}
In this section we will consider the Kratzer potential, which is written in the following form \cite{Berkdemir2005} \cite{Fues1926} \cite{Pliva1999}: 
\begin{equation}
V(r)=D_{e}\left( \frac{r-r_{e}}{r}\right) ^{2}  \label{41}
\end{equation}%
where the dissociation energy $D_{e}$ is the vertical distance between the dissociation limit and the minimum point of the potential curve, and $r_{e}$ is the inter-atomic separation at equilibrium. Using this expression in eq.(19), we obtain:
\begin{equation}
    \left[ \left( y\frac{d}{dr}\right) ^{2}+\frac{y^{2}}{r}\frac{d}{dr}-\frac{%
\left( l\left( l+1\right) +\frac{1}{4}+\frac{2mD_{e}r_{e}^{2}}{\hbar ^{2}}%
\right) y^{2}}{r^{2}} +\frac{4mD_{e}r_{e}}{\hbar ^{2}}\frac{y}{r}+\frac{2m\left(E-D_{e}\right) }{\hbar ^{2}}-\frac{\kappa \lambda }{2}\right] U\left(r\right) =0
\label{42}
\end{equation}

To solve this equation, we use the following change of variable: 
\begin{equation}
s=\frac{\sqrt{1+\kappa \lambda r^{2}}}{\sqrt{\kappa \lambda }r}  \label{43}
\end{equation}%
Thus, we can write it as follows:
\begin{equation}
\left[ \left( 1-s^{2}\right) ^{2}\frac{d^{2}}{ds^{2}}-s\left( 1-s^{2}\right) 
\frac{d}{ds}-\delta s^{2}+\eta s+\varepsilon \right] U(s)=0  \label{44}
\end{equation}%
with:%
\begin{equation}
\delta =\left( l+\frac{1}{2}\right) ^{2}+\frac{2mD_{e}r_{e}^{2}}{\hbar ^{2}}%
,\eta =\frac{4mD_{e}r_{e}}{\sqrt{\kappa \lambda }\hbar ^{2}}\text{ and }%
\varepsilon =\frac{2m\left( E-D_{e}\right) }{\kappa \lambda \hbar ^{2}}-%
\frac{1}{2}  \label{45}
\end{equation}

We can use the NU method, where the expressions of the polynomials appearing in eq.(7) are:%
\begin{equation}
\sigma \left( s\right) =1-s^{2},\ \tilde{\tau}\left( s\right) =-s\ \text{and 
}\tilde{\sigma}\left( s\right) =-\delta s^{2}-\eta s+\varepsilon  \label{46}
\end{equation}

If we insert them into eq.(14), we get:
\begin{equation}
\pi \left( s\right) =-\kappa \frac{s}{2}\pm \sqrt{\left( \frac{1}{4}+\delta
-\kappa k\right) s^{2}-\eta s+k-\varepsilon }  \label{47}
\end{equation}%
The constant $k$ is determined similarly to the PHO case. Therefore we get:%
\begin{equation}
\pi \left( s\right) =\left\{ 
\begin{array}{c}
\pi _{1,2}=\left( -\frac{\kappa }{2}\pm \zeta _{1}\right) s\mp \frac{\eta }{%
2\zeta _{1}},\text{\ for\ }k_{1}=\frac{1}{2}\left[ \varepsilon +\frac{\kappa 
}{4}+\kappa \delta +\kappa \sqrt{\vartriangle }\right]  \\ 
\pi _{3,4}=\left( -\frac{\kappa }{2}\pm \zeta _{2}\right) s\mp \frac{\eta }{%
2\zeta _{2}},\text{\ for\ }k_{2}=\frac{1}{2}\left[ \varepsilon +\frac{\kappa 
}{4}+\kappa \delta -\kappa \sqrt{\vartriangle }\right] 
\end{array}%
\right.   \label{48}
\end{equation}%
where:
\begin{equation}
\zeta _{1,2}=\sqrt{\frac{1}{4}+\delta -\kappa k_{1,2}}\text{ and}%
\vartriangle =\left( \varepsilon -\frac{\kappa }{4}-\kappa \delta \right)
^{2}-\eta ^{2}  \label{49}
\end{equation}%
In this case, the correct solution is $\pi _{1}$, so that:%
\begin{equation}
\tau \left( s\right) =2\left( \zeta _{1}-1\right) s-\frac{\eta }{\zeta _{1}}
\label{50}
\end{equation}%
From eq.(11), we have:%
\begin{equation}
\Lambda =k_{1}-\frac{\kappa }{2}+\kappa \zeta _{1}=n_{r}\left(
n_{r}+1-2\zeta _{1}\right) ,\text{ }n_{r}=0,1,2,...  \label{51}
\end{equation}%
Hence, the energy eigenvalues are:%
\begin{multline}
E_{n,l}=-\frac{4mD_{e}^{2}r_{e}^{2}}{2\hbar ^{2}}\left( n+\frac{1}{2}+\sqrt{%
\left( l+\frac{1}{2}\right) ^{2}+\frac{2mD_{e}r_{e}^{2}}{\hbar ^{2}}}\right)
^{-2}  \label{52} \\
-\frac{\kappa \lambda \hbar ^{2}}{2m}\left( \left( n+\frac{1}{2}+\sqrt{%
\left( l+\frac{1}{2}\right) ^{2}+\frac{2mD_{e}r_{e}^{2}}{\hbar ^{2}}}\right)
^{2}-\left( \left( l+\frac{1}{2}\right) ^{2}+\frac{2mD_{e}r_{e}^{2}}{\hbar
^{2}}\right) -\frac{3}{4}\right) 
\end{multline}%
In this situation, the deformation contributes evenly to energies, resulting in either increased or decreased binding energies depending on whether one is in the cases of dS or AdS, respectively.

To derive the complete expression of the wave functions $\psi \left(r,\theta ,\varphi \right) $ we use the relations of $\pi _{1}\left( s\right) $. First we get:%
\begin{equation}
\phi \left( s\right) =\left( 1+s\right) ^{^{\frac{1}{4}\left( 1-2\zeta _{1}-%
\frac{\eta }{\zeta _{1}}\right) }}\left( 1-s\right) ^{\frac{1}{4}\left(
1-2\zeta _{1}+\frac{\eta }{\zeta _{1}}\right) }  \label{53}
\end{equation}%
and using the Rodrigues formula, we find:%
\begin{equation}
y_{n}\left( s\right) =\frac{C_{n}}{\rho \left( s\right) }\frac{d^{n}}{ds^{n}}%
\left[ \left( 1-s^{2}\right) ^{n}\rho \left( s\right) \right]  \label{54}
\end{equation}%
\begin{equation}
\rho \left( s\right) =\left( 1+s\right) ^{^{\left( -\zeta _{1}-\frac{\eta }{%
2\zeta _{1}}\right) }}\left( 1-s\right) ^{^{\left( -\zeta _{1}+\frac{\eta }{%
2\zeta _{1}}\right) }}  \label{55}
\end{equation}

We see that eq.(54) stands for the Jacobi polynomials, so:
\begin{equation}
y_{n}\left( s\right) \equiv P_{n}^{\left( -\zeta _{1}-\frac{\eta }{2\zeta
_{1}},-\zeta _{1}+\frac{\eta }{2\zeta _{1}}\right) }\left( s\right)
\label{56}
\end{equation}%
Hence, $g(s)$ is written in the following form:%
\begin{equation}
U(s)=C_{n}\left( 1+s\right) ^{^{\frac{1}{4}\left( 1-2\zeta _{1}-\frac{\eta }{%
\zeta _{1}}\right) }}\left( 1-s\right) ^{\frac{1}{4}\left( 1-2\zeta _{1}+%
\frac{\eta }{\zeta _{1}}\right) }P_{n}^{\left( -\zeta _{1}-\frac{\eta }{%
2\zeta _{1}},-\zeta _{1}+\frac{\eta }{2\zeta _{1}}\right) }\left( s\right)
\label{57}
\end{equation}%
where $C_{n}$ is a normalization constant.

In terms of the variables $r,$ $\theta $\ and $\varphi $, we write the general form of the wave function $\Psi _{n}\left( r,\theta ,\varphi \right) $ as follows ($\chi _{\pm }=1-2\zeta _{1}\pm \frac{\eta }{\zeta _{1}}$):
\begin{equation}
    \Psi _{n}\left( r,\theta ,\varphi \right) =C_{n}r^{\frac{-1}{2}}\left( 1+\frac{\sqrt{1+\kappa \lambda r^{2}}}{\sqrt{\kappa \lambda }r}\right) ^{^{%
\frac{\chi _{-}}{4}}}\left( 1-\frac{\sqrt{1+\kappa \lambda r^{2}}}{\sqrt{\kappa \lambda }r}\right) ^{\frac{\chi _{+}}{4}} P_{n}^{\left( \frac{\chi _{-}-1}{2},\frac{\chi _{+}-1}{2}\right) }\left( \frac{\sqrt{1+\kappa \lambda r^{2}}}{\sqrt{\kappa \lambda }r}\right) 
 \label{58}
\end{equation}
In the case of anti-de Sitter space ($\kappa =-1$), the variable $s$ is imaginary and according to eq.(54), the solution must be multiplied by $\left( -i\right) ^{n}$. Here we can use the relation between the the complex Jacobi polynomials and the Romanovski polynomials:%
\begin{equation}
R_{n}^{\left( \alpha ,\beta \right) }\left( x\right) =\left( -i\right)
^{n}P_{n}^{\left( 1-\beta -\frac{i}{2}\alpha ,1-\beta +\frac{i}{2}\alpha
\right) }\left( ix\right)   \label{59}
\end{equation}%
to write the solutions in terms of Romanovsky polynomials ($\eta^{\backprime }=2ma/(\sqrt{\lambda }\hbar ^{2})$):%
\begin{equation}
\Psi _{n}\left( r,\theta ,\varphi \right) =C_{n}\lambda ^{\frac{1}{4}}\left( 
\sqrt{\lambda }r\right) ^{\zeta _{1}-1}e^{\frac{\eta ^{\backprime }}{2\zeta
_{1}}\tan ^{-1}\left( \frac{\sqrt{1-\lambda r^{2}}}{\sqrt{\lambda }r}\right)
}R_{n}^{\left( \zeta _{1}+1,\frac{\eta ^{\backprime }}{\zeta _{1}}\right)
}\left( \frac{\sqrt{1-\lambda r^{2}}}{\sqrt{\lambda }r}\right)   \label{60}
\end{equation}

\section{Discussions}
\label{sec:V}

In order to illustrate the impact of the deformed Heisenberg algebra leading to the extended uncertainty principle (EUP) on the bound states of PHO and Kratzer potentials in non-relativistic quantum mechanical systems, we have plotted the energy levels of s-states $E_{n,0}$ as a function of the
deformation parameter $\Lambda $ over different values of $n$.

As shown in eq.(52), the deformation contribution leads to a decrease in binding energy up to the ionization threshold in AdS space ($\kappa =-1$).
However, in dS space ($\kappa =+1$), deformation results in an increase in binding energy, and therefore, this critical behavior of ionization is not present. Eq.(52) is used to derive the expressions for the critical points corresponding to the values of the deformation parameter where $E_{n,l}=0$.%
\begin{equation}
\lambda _{c}\left( n,l\right) =\frac{m^{2}\left( n+\frac{1}{2}+\sqrt{\left(
l+\frac{1}{2}\right) ^{2}+\frac{2mb}{\hbar ^{2}}}\right) ^{-2}}{\hbar
^{4}\left( n+\frac{1}{2}+\sqrt{\left( l+\frac{1}{2}\right) ^{2}+\frac{2mb}{%
\hbar ^{2}}}\right) ^{2}-\left( \left( l+\frac{1}{2}\right) ^{2}+\frac{2mb}{%
\hbar ^{2}}\right) -\frac{3}{4}}  \label{61}
\end{equation}

We use the Hartree atomic units $\ \hbar =c=4\pi \varepsilon _{0}$ and the values of the potential parameters are taken from \cite{Oyewumi2012}.
\begin{table}[H]
	\caption{The potential parameters \cite{Oyewumi2012}}
	\centering
\begin{tabular}{|c|c|c|c|}
\hline
Parameters & $N_{2}$ & $H_{2}$ & $CO$ \\ \hline
$D_{e}$($eV$) & $11,9382$ & $4,7446$ & $10.8451$ \\ \hline
$r_{e}$($A^{\circ }$) & $1,0940$ & $0,7416$ & $1.1283$ \\ \hline
$m$($amu$) & $7,00335$ & $0,50391$ & $6.86059$ \\ \hline
\end{tabular}
\label{tab:table.1}
\end{table}
Table.2 shows the critical values when the Kratzer potential model is used for different diatomic molecules such as $N_{2}$, $H_{2}$ and $CO$.

\begin{table}[H]
	\caption{Critical values of $\Lambda$ for $E_{n,l}=0$ in the case of the Kratzer potential}
	\centering
\begin{tabular}{|c|c|c|c|c|}
\hline
$n$ & $l$ & $N_{2}$ & $H_{2}$ & $CO$ \\ \hline
$1$ & $0$ & $0,19770$ & $0,00220$ & $0,17762$ \\ \hline
$2$ & $0$ & $0,08419$ & $0,00052$ & $0,07507$ \\ \hline
& $1$ & $0,01529$ & $0,00023$ & $0,06919$ \\ \hline
$3$ & $0$ & $0,043963$ & $0,000179$ & $0,03897$ \\ \hline
& $1$ & $0,04970$ & $0,000095$ & $0,03621$ \\ \hline
& $2$ & $0,03599$ & $0,000052$ & $0,03164$ \\ \hline
$4$ & $0$ & $0,022572$ & $0,000077$ & $0,02269$ \\ \hline
& $1$ & $0,02412$ & $0,000045$ & $0,02122$ \\ \hline
& $2$ & $0,021423$ & $0,000027$ & $0,01876$ \\ \hline
& $3$ & $0,018307$ & $0,000017$ & $0,01595$ \\ \hline
$5$ & $0$ & $0,016227$ & $0,000038$ & $0,01426$ \\ \hline
& $1$ & $0,015294$ & $0,000024$ & $0,01341$ \\ \hline
& $2$ & $0,013718$ & $0,000015$ & $0,01197$ \\ \hline
& $3$ & $0,011856$ & $0,000010$ & $0,01030$ \\ \hline
& $4$ & $0,01007$ & $7,4\times10^{-6}$ & $0,00866$ \\ \hline
\end{tabular}
\label{tab:table.2}
\end{table}
\begin{figure} [H]
    \centering
    \includegraphics[scale=0.7]{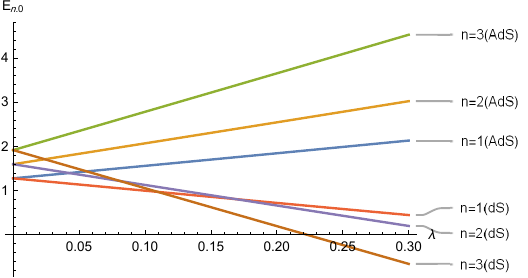}
    \caption{PHO Energies for $CO$ in both dS and AdS spaces}
    \label{fig:2}
\end{figure}

\begin{figure} [H]
    \centering
    \includegraphics[scale=0.7]{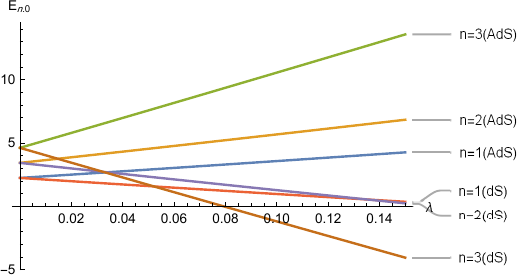}
    \caption{PHO Energies for $H_{2}$ in both dS and AdS spaces}
    \label{fig:3}
\end{figure}

\begin{figure} [H]
    \centering
    \includegraphics[scale=0.7]{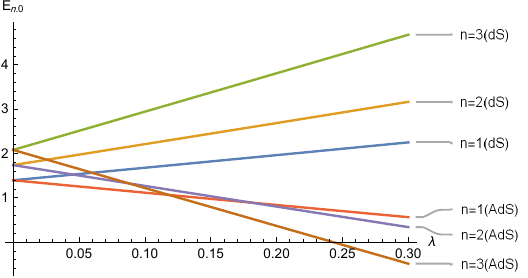}
    \caption{PHO Energies for $N_{2}$ in both dS and AdS spaces}
    \label{fig:4}
\end{figure}
The deformation contribution can alter the order of energy levels, thus redefining the fundamental level. Tables 3 and 4 provide the numerical values of these values $\lambda _{f}\left( n,l\right) $ responsible for the inversion between higher energy levels and the ordinary fundamental level.
\begin{table}[H]
	\caption{Values of  $\lambda _{f}\left( n,l\right) $ for the PHO case}
	\centering
\begin{tabular}{|c|c|c|c|c|}
\hline
$n$ & $l$ & $N_{2}$ & $H_{2}$ & $CO$ \\ \hline
$1$ & $0$ & $0.0597119$ & $0.00162457$ & $0.0567986$ \\ \hline
$2$ & $0$ & $0.0352453$ & $0.000503812$ & $0.0332321$ \\ \hline
& $1$ & $0.0379421$ & $0.000300942$ & $0.0354058$ \\ \hline
$3$ & $0$ & $0.0221267$ & $0.00020685$ & $0.0207211$ \\ \hline
& $1$ & $0.0236383$ & $0.00012418$ & $0.0219046$ \\ \hline
& $2$ & $0.0254385$ & $0.00009019$ & $0.0232438$ \\ \hline
$4$ & $0$ & $0.0145841$ & $0.00010054$ & $0.0135836$ \\ \hline
& $1$ & $0.0154871$ & $0.00006065$ & $0.0142719$ \\ \hline
& $2$ & $0.0165328$ & $0.0000445466$ & $0,177242$ \\ \hline
& $3$ & $0.0171201$ & $0.0000361096$ & $0.0153659$ \\ \hline
$5$ & $0$ & $0.00999948$ & $0.0000547581$ & $0.0092722$ \\ \hline
& $1$ & $0.0105673$ & $0.0000331766$ & $0.0096943$ \\ \hline
& $2$ & $0.0112105$ & $0.0000245916$ & $0.0101414$ \\ \hline
& $3$ & $0.0115529$ & $0.0000201296$ & $0.0103248$ \\ \hline
& $4$ & $0.0115344$ & $0.000017249$ & $0.0102143$ \\ \hline
\end{tabular}
\label{tab:table.3}
\end{table}
\begin{table}[H]
	\caption{Values of  $\lambda _{f}\left( n,l\right) $ for the Kratzer case}
	\centering
\begin{tabular}{|c|c|c|c|c|}
\hline
$n$ & $l$ & $N_{2}$ & $H_{2}$ & $CO$ \\ \hline
$0$ & $0$ &  &  &  \\ \hline
$1$ & $0$ & $0,847672$ & $0,211971$ & $0.775011$ \\ \hline
& $1$ & $0.772393$ & $0.159601$ & $0.704656$ \\ \hline
$2$ & $0$ & $0,489404$ & $0,122382$ & $0.447453$ \\ \hline
& $1$ & $0,448771$ & $0,0936605$ & $0.40947$ \\ \hline
& $2$ & $0.399568$ & $0.0821475$ & $0.364012$ \\ \hline
$3$ & $0$ & $0,346061$ & $0,0865369$ & $0.316397$ \\ \hline
& $1$ & $0,317836$ & $0,0665504$ & $0.290011$ \\ \hline
& $2$ & $0,283484$ & $0,0584089$ & $0.25827$ \\ \hline
& $3$ & $0.255954$ & $0.0543813$ & $0.233107$ \\ \hline
$4$ & $0$ & $0,268058$ & $0,0670312$ & $0.24508$ \\ \hline
& $1$ & $0,246352$ & $0,051601$ & $0.224788$ \\ \hline
& $2$ & $0,219881$ & $0,0453442$ & $0.200328$ \\ \hline
& $3$ & $0,198629$ & $0,042228$ & $0.180902$ \\ \hline
& $4$ & $0.183597$ & $0.0403857$ & $0.167265$ \\ \hline
$5$ & $0$ & $0,218868$ & $0,0547307$ & $0.200107$ \\ \hline
& $1$ & $0,20121$ & $0,0421667$ & $0.183599$ \\ \hline
& $2$ & $0,179653$ & $0,0370648$ & $0.163679$ \\ \hline
& $3$ & $0,16233$ & $0,0345219$ & $0.147844$ \\ \hline
& $4$ & $0,1500$ & $0,0330181$ & $0.136721$ \\ \hline
& $5$ & $0.141576$ & $0.0320283$ & $0.129049$ \\ \hline
\end{tabular}%
\label{tab:table.4}
\end{table}

\begin{figure} [H]
    \centering
    \includegraphics[scale=0.7]{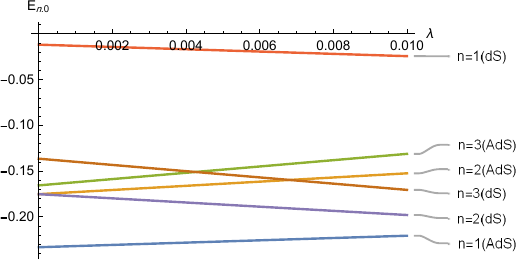}
    \caption{Kratzer Energies for $CO$ in both dS and AdS spaces}
    \label{fig:5}
\end{figure}

\begin{figure} [H]
    \centering
    \includegraphics[scale=0.7]{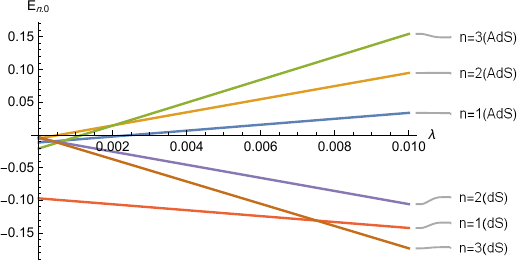}
    \caption{Kratzer Energies for $H_{2}$ in both dS and AdS spaces}
    \label{fig:6}
\end{figure}

\begin{figure} [H]
    \centering
    \includegraphics[scale=0.7]{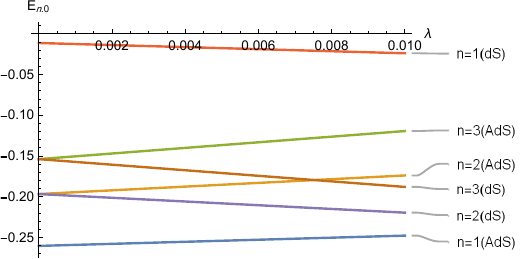}
    \caption{Kratzer Energies for $N_{2}$ in both dS and AdS spaces}
    \label{fig:7}
\end{figure}

\section{Conclusion}
\label{sec:VI}
This work presents an analytical study of the three-dimensional Schr\"{o}dinger equation in deSitter and anti-deSitter spaces for the pseudoharmonic oscillator and Kratzer potential. The position representation of the extended uncertainty principle formulation of the deformed spaces was
used, and the Nikiforov--Uvarov method was applied to compute the exact expressions of the eigenenergies and eigenfunctions.

Regarding the radial part of the eigenfunctions, it corresponds to the Jacobi polynomials for the PHO potential and the Romanovski polynomials for the Kratzer potential.

In the PHO case, the contribution of deformation to energy eigenvalues is not uniform and varies from one level to another. These variations can even cause a reversal in the level order for extreme values of the parameter $\Lambda$, potentially redefining the fundamental level of the system if the inversion is between a higher energy level and the ordinary fundamental level ($E_{n,l}<E_{0,0}$).

The Kratzer case shows that deformation contributes evenly to energies, resulting in either increased or decreased binding energies depending on whether one is in dS or AdS space. In AdS case, the decrease of the binding energy can lead the ionization of the system ($E_{n,l}=0$). However, in dS case, deformation results in an increase in binding energies and therefore, this critical behavior of ionization is absent.

For the PHO case, critical values of the deformation parameter have been computed for the inversion between an excited level and the fundamental one in different diatomic molecules, including $N_{2}$, $H_{2}$, and $CO$. The critical values for ionisation in the Kratzer case were also determined for these diatomic molecules.

\end{document}